# Effect of transition radiation and dispersion spreading at achievement of a large transformer ratio of driver-bunch energy into wakefield in a dielectric resonator accelerator

*Vasyl Maslov (1, 2), Denys Bondar (1, 2), Ivan Onishchenko (1), Valery Papkovich (1, 2)*
*((1) National Science Center "Kharkov Institute of Physics and Technology", Kharkov, Ukraine*
*(2) V.N. Karazin Kharkov National University, Kharkov, Ukraine)*

vmaslov@kipt.kharkov.ua    bondar@kipt.kharkov.ua

Possibility of increase of transformer ratio TR in the case of the profiled sequence of bunches at their injection in a two-beam electron-positron dielectric-resonator collider is considered. Unlike considered earlier the waveguide case, for which TR is equal to the doubled number of bunches of sequence which provide a contribution to the total wakefield, and which is limited by the effect of the group velocity, in a resonator this limitation is absent. For derivation in the case of resonator of TR, proportional to the number of bunches, as well as in the waveguide, the ratio of bunch charges is selected to be equal 1:3:5: … , lengths of bunches, equal to the half of wavelength, interval between bunches is multiple to the wavelength. The effect of transition radiation and dispersion spreading on the transformer ratio is studied by numerical simulation.

## INTRODUCTION

It is useful to use a resonator instead of a waveguide as an accelerating structure in a multi-bunch scheme of a wakefield accelerator, for which a large charge driver is replaced by a resonant sequence of moderate charge driver-bunches with an equivalent total charge. In the case of a waveguide, as shown earlier, due to the removal of the excited wakefield at the output of the waveguide, the number of bunches, whose wakefields contribute to the total wakefield, is limited by the length of the waveguide and in real experiments is small. In the resonator case, due to feedback in the form of reflection of the excited field from the ends of the resonator (equivalent to the long waveguide), this quantity is limited only by the Q-factor of the resonator.

In a resonator, the wakefield, excited by bunches, after the release of bunches from a resonator has the form of pulses, which oscillate with the group velocity $V_g$ between the ends of the resonator. The excitement of a wakefield in the form of pulses is an advantage, as it contributes to an increase in the threshold of breakdown. While the 1-st bunch is in the resonator, the front edge of the pulse has a velocity $V_b$, and the rear is the group velocity $V_g$. In the resonance excitation of the wakefield by successive bunches at moments of time when the back front of the pulse is at the injection boundary of resonator, on the back front of the pulse there is a decelerating for electrons wakefield and a wakefield accelerating for the positrons. At moments of time when the forward front of the pulse is detected at the end of the resonator at the front of the pulse there is also the maximum decelerating field for electrons, and accelerating field for positrons. Therefore, in the case of accelerating electrons in the resonator, it is advantageous to use the case of charge-shaped bunches, since in this case a large transformer ratio of the energy of the driver-bunches into the energy of witness-bunches is achieved, as well as at the moment of injection of witness-bunches on the boundary of injection, the back front of the pulse with the accelerating phase of the wakefield. In the case of accelerating the positrons in the resonator, it is advantageous to use the case of resonance excitation of the wakefield by successive bunches, since in this case, at the moment of injection of accelerated positrons at the injection boundary the back front of the pulse with the accelerating phase of the field is detected.

We consider the injection of driver-bunches, rectangular cylinders, whose length is equal to the half-wavelength $\xi_b=\lambda/2$, in the dielectric resonator. In this paper we consider the problem of increasing the transformer ratio (according to limitation TR≤2 of the Wilson theorem in a waveguide case) for a charge-shaped sequence of bunches, but for a resonator case. The maximum energy to which electrons/positrons can be accelerated at some energy of the electron driver-bunches of sequence, which excite wakefield in two-beam electron-positron dielectric resonator collider, is determined by the transformer ratio (see [1-35]). The transformer ratio, defined as ratio $TR=\dfrac{E_2}{E_1}$ of the wakefield $E_2$, which is excited in dielectric resonator accelerator by sequence of the electron bunches, to the field $E_1$, in which an electron bunch is decelerated, is considered with charge shaping of rectangular in longitudinal direction (current is const along each bunch) bunches according to linear law along sequence [3, 4], so that ratio of charges of bunches of sequence equals 1:3:5: … [3, 4]. The bunch length equals to half of wave-length $\Delta\xi_b=\lambda/2$. The choice of such length of bunches is determined by the necessity to provide not only large TR but high accelerating gradient (wakefield) too excited by sequence of N bunches and by the necessity to damp the transition radiation. The latter is determined by the following. Since when a large TR is obtained, a comparatively small wakefield, decelerating bunches-drivers, is inevitable, therefore, the transition radiation can strongly distort a small decelerating wakefield and, therefore, strongly distort TR.

The porosity between bunches is multiple of wavelength $\delta\xi=p\lambda$, p=1, 2, ... A next bunch is injected in the resonator, when the back wavefront of wakefield pulse, excited by previous bunches, is on the injection boundary ($z = 0$). A next bunch leaves the resonator, when the first wavefront of wakefield pulse, excited by previous bunches, is on the end of the resonator. Then wakefield pulses, excited by all consistently injected bunches, are coherently added. In other words, coherent accumulation of wakefield is realized. The conditions have been formulated, when decelerating longitudinal wakefield for all bunches is small, identical but inhomogeneous along their length. Then one can provide a large transformer ratio TR. Several conditions should be satisfied for this purpose. The wakefield and transformer ratio have been derived after N-th bunch.

Also the computer simulation of wakefield excitation in the cylindrical dielectric resonator by the train of three bunches has been performed. The contribution of dispersion spreading and transition radiation (see [20]), excited at the input and output boundaries of the resonator, into the wakefield is taken into account. It has been shown that the dependence R=2N, which follow from the theory for Cherenkov radiation is performed approximately.

Advantage of wakefield accelerator with large TR that when accelerated electron bunch is injected in the resonator, the back wavefront of wakefield pulse and maximum accelerating field, excited by decelerated electron bunches, is on the injection boundary.

For effective acceleration of positron bunches it is necessary to use the resonant excitation of wakefield. Then when accelerated positron bunch is injected in the resonator, the back wavefront of wakefield pulse and maximum accelerating field, excited by decelerated electron bunches, is on the injection boundary.

## 1. TRANSFORMER RATIO AT WAKEFIELD EXCITATION IN DIELECTRIC RESONATOR BY SEQUENCE OF RECTANGULAR ELECTRON BUNCHES WITH LINEAR GROWTH OF CHARGE

In this paper the transformer ratio TR is investigated theoretically. In many cases transformer ratio can be concluded to the ratio of maximum accelerating wakefield, experienced by witness bunch, to the maximum decelerating wakefield, experienced by driver bunches.

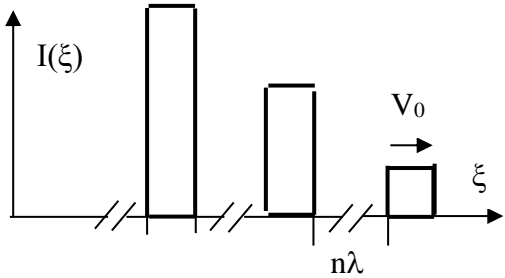

*Fig. 1. The current distribution of sequence of rectangular bunches, charge of which is shaped*

In [18] the expression for the wakefield, excited in a dielectric resonator by the sequence of electron bunches, each of which is a infinitely thin ring, has been derived.

So the charge density of sequence of rectangular bunches is distributed according to (see Fig. 1)

$$n_b(z,t) = n_{b0}(2N-1), \quad N \geq 1, \quad 0 < V_0\left(t-T(N-1)\right)-z < \Delta\xi_b,$$

$$T(N-1) < t < T(N-1) + \frac{(L+\Delta\xi_b)}{V_0}. \quad (1)$$

Then the ratio of charges $Q_N$ of consecutive bunches equals to known values 1:3:5 …

A next N+1-th bunch is injected in the resonator, when the back wavefront of wakefield pulse, excited by previous N bunches, is on the injection boundary ($z = 0$) (see Fig. 2).

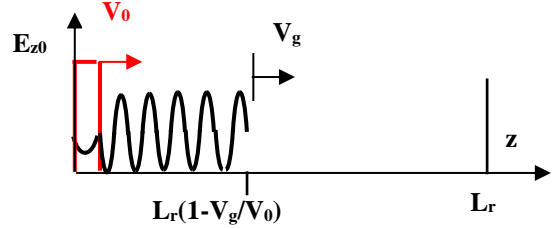

*Fig. 2. A schematic of the wakefield pulse, excited by previous N bunches, when N+1-th bunch is injected in the resonator*

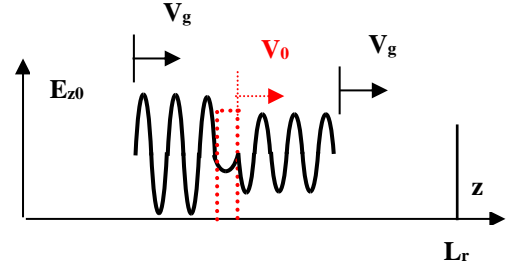

*Fig. 3. An approximate view of the wakefield pulse, excited by previous N bunches and excited by N+1-th bunch, when N+1-th bunch is in the middle of the resonator*

An approximate view of the wakefield pulse, excited by previous N bunches and excited by N+1-th bunch, when N+1-th bunch is in the middle of the resonator, is shown in Fig. 3. Excited longitudinal decelerating wakefield $E_z$ is small and identical for all bunches but non-uniform along them. Then one can provide a large transformation ratio R.

A next N+1-th bunch leaves the resonator, when the first wavefront of wakefield pulse, excited by N+1 bunches, is on the end of the resonator (z = L) (see Fig. 4).

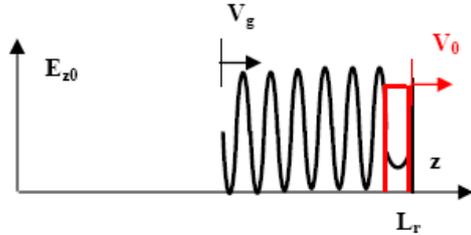

*Fig. 4. A schematic of the wakefield pulse, excited by N+1 bunches, when N+1-th bunch leaves the resonator*

For achieving a large transformation ratio R several conditions should be satisfied. Namely, we choose the length of the resonator L, the group velocity $V_g$, the bunch repetition frequency $\omega_m$ and the wave frequency, which satisfy the following equalities

$$T = \frac{2L_r}{V_g} = \frac{2\pi}{\omega_m} = \frac{\pi q}{\omega_0}, \quad q = 1, 3, ..., \quad \frac{V_g}{V_0} = \frac{4L_r}{q\lambda} \quad (2)$$

Then for the selected length of the resonator and q, equal $\frac{L}{\lambda} = 4$ and $q = 20$ group velocity should be equal $\frac{V_g}{V_0} = 0.8$. $V_0$ is the beam velocity. For $\frac{L}{\lambda} = 5$ and $q = 32$ group velocity should be equal $\frac{V_g}{V_0} = 0.625$.

Thus, all the next bunches after the first one begin to be injected in the resonator (on the boundary $z = 0$), when the trailing edge of the wakefield pulse, created by the previous bunches, is located at the point $z = 0$. At this moment the leading edge of the wakefield pulse, located at the distance from the injection boundary, equal to $L\left(1 - \frac{V_g}{V_0}\right) + \Delta\xi_b$ (see Fig. 2), is located at the distance $L\left(\frac{V_g}{V_0}\right) - \Delta\xi_b$ from the end of the resonator ($z = L$).

Again injected bunch reaches the end of the resonator together with the leading edge of the wakefield pulse, created by the previous bunches. Then wakefield pulses, excited by all consistently injected bunches, are coherently added. In other words, coherent accumulation of wakefield is realized.

At wakefield pulse excitation by the 1-st bunch the wakefield in the whole resonator (one can derive, using [16, 18]) within the time $0 < t < \frac{L_r + \Delta\xi_b}{V_0}$ is proportional to

$$Z_\parallel(z, t) = \left(\frac{1}{k}\right)\left[\theta(V_0 t - z) - \theta(V_0 t - \Delta\xi_b - z)\right] \sin\left[k(V_0 t - z)\right] + \\ + \left(\frac{2}{k}\right)\left[\theta(V_0 t - \Delta\xi_b - z) - \theta(V_g t - z)\right] \sin\left[k(V_0 t - z)\right]. \quad (3)$$

The 1-st term is the field inside of the 1-st bunch, the 2nd term is the wakefield after the 1st bunch. Thus, after 1-st bunch $T_E = 2$.

Inside the 2-nd bunch $0 < \xi = V_0(t - T) - z < \Delta\xi_b$ the wakefield on the times $T < t < T + \frac{L_r + \Delta\xi_b}{V_0}$ is proportional to

$$Z_\parallel(z, t) = \quad (4)$$
$$\left[\theta(V_0(t-T) - z) - \theta(V_0(t-T) - \Delta\xi_b - z)\right] k^{-1} \sin\left[k(V_0(t-T) - z)\right].$$

The decelerating field into the 2-nd bunch equals to decelerating field into the 1-st bunch.

After the 2-nd bunch $\xi = V_0(t-T) - z > \Delta\xi_b$ on the times $T < t < T + (L_r + \Delta\xi_b)/V_0$ it excites wakefield, which is proportional to

$$Z_\parallel(z, t) = \left[\theta(V_0(t-T) - \Delta\xi_b - z) - \theta(V_g(t-T) - z)\right] \times \\ \times 4k^{-1} \sin\left[k(V_0(t-T) - z)\right]. \quad (5)$$

Thus, after 2-nd bunch $T_E = 4$.

At wakefield pulse excitation by the N-th bunch the wakefield in whole resonator within the time $T(N-1) \leq t \leq T(N-1) + \frac{(L + \Delta\xi_b)}{V_0}$, $T = \frac{2L}{V_g}$, is proportional to

$$Z_\parallel(z, t) = \quad (6)$$
$$\left(\frac{1}{k}\right)\left[\theta(V_0(t-T(N-1)) - z) - \theta(V_0(t-T(N-1)) - \Delta\xi_b - z)\right] \sin(k\xi)$$
$$+ \begin{bmatrix} \theta(V_0(t-T(N-1)) - \Delta\xi_b - z) - \theta(V_g(t-T(N-1)) - z) \end{bmatrix} \times \\ \times \left(\frac{2N}{k}\right) \sin(k\xi) \quad +$$
$$+ \begin{bmatrix} \theta(V_g(t-T(N-1)) + L_r\left(1 - \frac{V_g}{V_0}\right) - z) - \theta(V_0(t-T(N-1)) - z) \end{bmatrix} \times \\ \times \left(\frac{2}{k}\right)(N-1) \sin(k\xi).$$

Here $\xi \equiv V_0 t - z$, the 1-st term is the decelerating field inside the N-th bunch, the second term is the wakefield after the N-th bunch, the 3-rd term is the field before the N-th bunch, excited by N-1 bunches. Thus, after the N-th bunch the transformation ratio is equal to $T_E = 2N$, similar to [1, 3, 4]. The decelerating field inside the N-th bunch is equal to the decelerating field inside the 1-st bunch.

## 2. THE NUMERICAL SIMULATION OF THE EFFECT OF THE DISPERSION SPREADING AND CONTRIBUTION OF TRANSITION RADIATION IN WAKEFIELD

Now we considered the effect of the dispersion spreading and contribution of transition radiation in wakefield, which is excited by a sequence of electron bunches in resonator.

The dependence of the ratio of amplitudes of Cerenkov and transition radiation, which are excited by bunch in a cylindrical dielectric resonator, on the number of radial modes has been considered for: $\varepsilon = 3$, $b = 0.5$ cm, $L = 3$ cm. The radius of bunch equals $r_b = 0.015$ cm. $\varepsilon$ is the dielectric permeability, b is the inner radius of the metal tube. When the bunch almost reaches the end of resonator the distribution of $E_z$ presented in Fig. 1. The perturbation is two pulses: 1-st is after the bunch, 2-nd pulse is near the front, moving with $V_g$. During wakefield oscillations between the ends the distribution of $E_z$ varies because $V_{ph}$ and $V_g$ are different.

In the case of a single radial mode the amplitude ratio of Cerenkov and transition radiation equals 1.5. When the number of excited radial modes equals 6, this ratio equals 10. Such small contribution of transition radiation to wakefield is very important for controlled wakefield excitation with large TR.

Reduction of the contribution of transition radiation in wakefield due to growth of the number of modes is determined that the Cerenkov peaks of different modes are coherently added. Pulses of transition radiation are very variable with the growth of number of excited modes (Fig. 5). This results that transition radiation pulses are added incoherent.

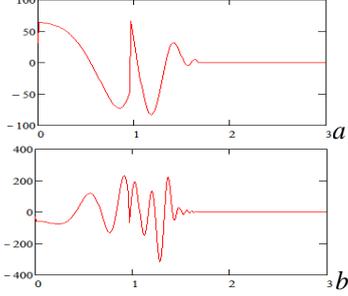

*Fig. 5. Two-mode pulse of transition radiation, which is excited by point bunch in a dielectric resonator (a); Six-mode pulse of transition radiation, which is excited by point bunch in a dielectric resonator (b)*

We consider the effect of dispersion spreading on the example of wakefield excitation in the resonator by the resonant ($\omega = h\omega_m$, h is the integer) sequence. N+1-th bunch is injected into resonator when the trailing edge of pulse is on the injection boundary.

We choose L, $\varepsilon$, $\omega_m$ and $\lambda$, which satisfy
$$T = 2L\beta\varepsilon/c = 2\pi/\omega_m, \quad 2L(\beta^2\varepsilon - 1) = n\lambda, \quad (7)$$

n is the integer. Then for L=0.5473 cm, b=0.025 cm, $\varepsilon$=1.7 we have n=14, L/$\lambda$=10. The distribution of pulse, which is excited by N+1-th bunch, is shown in Fig. 6.

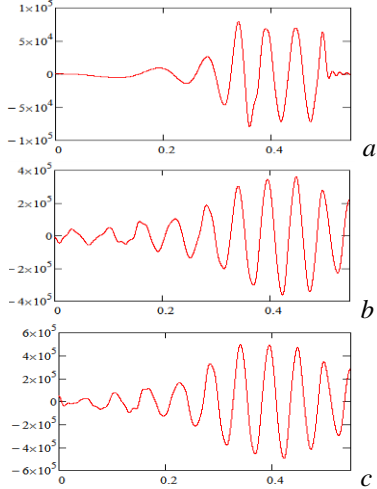

*Fig. 6. Wakefield pulse, which is excited by 1-st bunch in the dielectric resonator (a); Wakefield pulse, which is excited by 5-th bunch in the dielectric resonator (b); Wakefield pulse, which is excited by 8-th bunch in the dielectric resonator (c)*

Dispersion spreading and transition radiation lead to the fact that in the neighborhood of Cerenkov pulse precursor and tail are formed. They do not destruct pulse so that after injection of 10 bunches the wakefield amplitude is increased about 10 times in the presence of background perturbations.

The ratio $E_z^{(c)}/E_z^{(t)}$ of amplitudes of CR $E_z^{(c)}$ and TrR $E_z^{(t)}$ has been considered, which are excited in a dielectric resonator. It is shown that for b=0.025 cm, $r_b$=0.015 cm, $\varepsilon$=1,7 it equals (Fig. 7) $E_z^{(c)}/E_z^{(t)} \approx 2.5$.

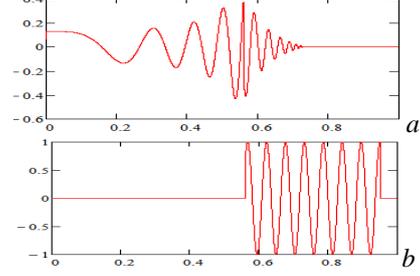

*Fig. 7. Pulse of transition radiation which is excited in resonator by point bunch (a); Cerenkov pulse which is excited in resonator by point bunch (b)*

From comparison of pulses, excited by a bunch-disk and by a bunch with $\xi_b=\lambda/2$, (Fig. 8) one can see that in the last case WF amplitude is more uniform.

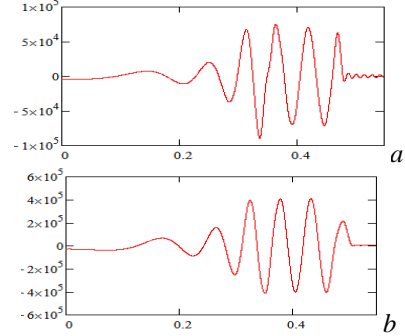

*Fig. 8. Wakefield pulse, excited in dielectric resonator by a bunch-disk (a); Wakefield pulse, excited in dielectric resonator by a bunch with $\xi_b=\lambda/2$ (b)*

The reason for this is clear from comparison of transition radiation, excited by a point bunch and by a bunch with $\xi_b=\lambda/2$ (Fig. 9).

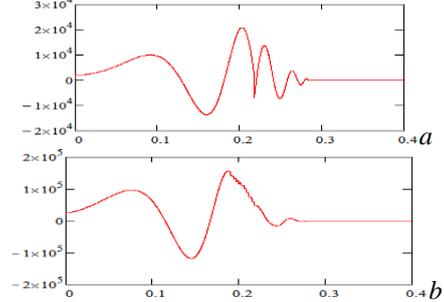

*Fig. 9. Pulse of transition radiation, excited in dielectric resonator by a point bunch (a); Fig. 9. Pulse of transition radiation, excited in dielectric resonator by a bunch with $\xi_b=\lambda/2$ (b)*

At bunch length increase HF parts of transition radiation, which are excited by different layers of bunch, are added incoherently, because they are very quick-changeable.

## 3. THE RESULTS OF NUMERICAL SIMULATION ON TRANSFORMER RATIO

The main purpose of the numerical simulation of the wakefield excitation in a cylindrical dielectric resonator with the parameters: $\varepsilon$=1.725, b=0.25 mm, Q=2 nC, L=5.56 mm, $\lambda$=0.556 μm, $\omega/2\pi$=539.6 GHz, $\omega_m/2\pi$=15.6397 GHz by the sequence of electron bunches $\varepsilon_b=1$ GeV with radius $r_b=0.15$mm with a Gaussian radial distribution, is a demonstration of the achievement of a large transformer ratio TR, which is defined as the ratio of the maximum amplitude of the accelerating wakefield $E_{acc}$ after the driver-bunch to the maximum decelerating wakefield $E_{dec}$ in the middle of the driver-bunch $TR=-E_{acc}/E_{dec}$. In Figs. 1-3 show the samples of the excited wakefield in the resonator and the location of the rear driver-bunch of sequence (upper arrow) and of accelerated bunch (lower arrow) for cases of one bunch (Fig. 10) of two bunches (Fig. 11) and ten bunches (Fig. 12) in a sequence when they are at the distance z = 0.15 cm from the injection boundary.

The influence of dispersion spreading and transition radiation, excited at the input and output boundaries of the resonator, on the transformer ratio enhancement is taking into account in computer simulation (see Figs. 10-12), using the theory, developed in [36].

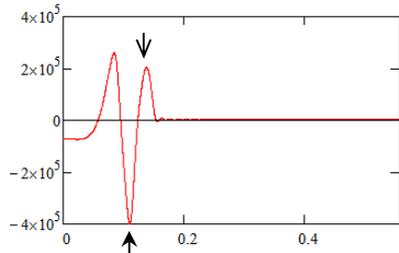
*Fig. 10. $E_z$ (in stavolt/cm), excited by 1-st bunch, when it is at the distance z=0.15 cm from the injection boundary*

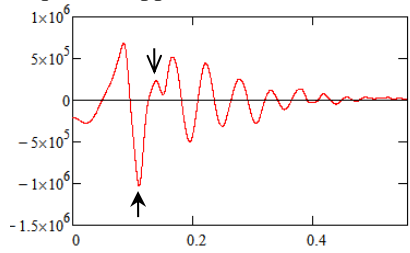
*Fig. 11. $E_z$, excited by two bunches, when $2^{nd}$ bunch is at the distance z=0.15 cm from the injection boundary*

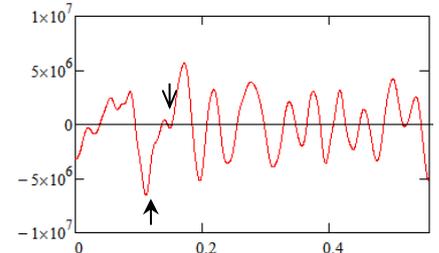
*Fig. 12. $E_z$, excited by ten bunches, when $10^{th}$ bunch is at the distance z=0.15 cm from the injection boundary*

Figs. 10-12 show that in the case of a single bunch (Fig. 10), the transformer ratio TR=2, in accordance with Wilson's theorem, while with increasing number of bunches of a sequence with a linearly increasing charge (in the presence of a frequency detuning of the excited wakefield frequency with a frequency of following bunches $\omega/\omega_m=34.5$), the transformer ratio TR increases. For cases of two bunches, TR=4.4 (Fig. 11) and ten bunches TR=20 (Fig. 12). Consequently, the transformer ratio TR increases in direct proportion to the number of exciting bunches.

Taking into account the excited transition radiation does not change essentially TR value.

However, the ratio of the accelerating wakefield $E_{acc}$ for a witness-bunch to the maximum decelerating wakefield $E_{dec}$ in the middle of the driver bunch varies greatly along the resonator. The longitudinal fields $E_z$ (in stavolt/cm) after the 2nd bunch, counted at 18 points along the resonator from z = 0.1 cm to z=0.525 cm, are given in the table below and in Fig. 13.

| z = 0.1 | z = 0.125 | z = 0.15 | z = 0.175 | z = 0.2 | z= 0.225 | z= 0.25 | z= 0.275 | z= 0.3 |
|---|---|---|---|---|---|---|---|---|
| $E_{dec}=$ 3.5×10$^5$ | $E_{dec}=$ 4×10$^5$ | $E_{dec}=$ 2.5×10$^5$ | $E_{dec}=$ 2.5×10$^5$ | $E_{dec}=$ 2×10$^5$ | $E_{dec}=10^5$ | $E_{dec}=$ 1.2×10$^5$ | $E_{dec}=10^5$ | $E_{dec}=10^5$ |
| $E_{ac}=$ -4×10$^5$ | $E_{ac}=$ -3.2×10$^5$ | $E_{ac}=$ -9.2×10$^5$ | $E_{ac}=$ -7.8×10$^5$ | $E_{ac}=$ -8.8×10$^5$ | $E_{ac}=$ -1.2×10$^6$ | $E_{ac}=$ -1.3×10$^6$ | $E_{ac}=$ -1.2×10$^6$ | $E_{ac}=$ -1.2×10$^6$ |
| z= 0.325 | z= 0.35 | z= 0.375 | z= 0.4 | z= 0.425 | z= 0.45 | z= 0.475 | z= 0.5 | z= 0.525 |
| $E_{dec}=10^5$ | $E_{dec}=$ 1.3×10$^5$ | $E_{dec}=$ 2,1×10$^5$ | $E_{dec}=$ 2,4×10$^5$ | $E_{dec}=$ 3.2×10$^5$ | $E_{dec}=$ 3.1×10$^5$ | $E_{dec}=$ 3.1×10$^5$ | $E_{dec}=$ 4.4×10$^5$ | $E_{dec}=$ 4.7×10$^5$ |
| $E_{ac}=$ -1.2×10$^6$ | $E_{ac}=$ -1.1×10$^6$ | $E_{ac}=$ -9.2×10$^5$ | $E_{ac}=$ -8.3×10$^5$ | $E_{ac}=$ -6.1×10$^5$ | $E_{ac}=$ -6.7×10$^5$ | $E_{ac}=$ -6.5×10$^5$ | $E_{ac}=$ -4.6×10$^5$ | $E_{ac}=$ -2.7×10$^5$ |
| **Table 1** | | | | | | | | |

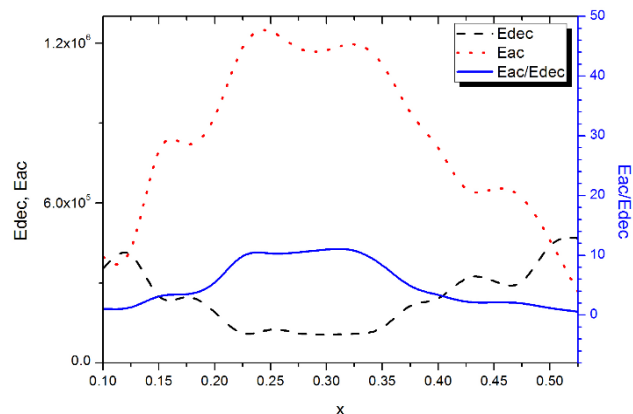
*Fig. 13. The decelerating field (in stavolt/cm) inside the 2nd bunch $E_{dec}$, $E_{ac}=-E_{acc}$, $E_{acc}$ is the accelerating field after the 2nd bunch and their ratio $E_{ac}/E_{dec}$ along the resonator*

One can see that the ratio of the accelerating field to the decelerating field is about 1 near the ends. In the middle of the resonator, this ratio reaches 11. The ratio of the acquired energy on the length of the resonator to the lost energy on the length of the resonator, calculated at 18 points, is equal to TR=3.4 after the 2nd bunch. This is less than the theoretical value TR=2N=4, but more than 1.5N. N is the number of injected bunches. In this case, the decelerating field is minimal in the middle of the resonator and grows to its ends. The accelerating field, on the contrary, is maximal in the middle of the resonator and decreases to its ends.

After 3rd bunch, the ratio of the acquired energy on the length of the resonator to the lost energy on the length of the resonator is equal to TR=4.8≈1.5N.

After 4th bunch, the ratio of the acquired energy on the length of the resonator to the lost energy on the length of the resonator is equal to N<TR=5<1.5N.

After 5th bunch, the ratio of the acquired energy on the length of the resonator to the lost energy on the length of the resonator is equal to N<TR=6.2<1.5N.

After 9th bunch, the ratio of the acquired energy on the length of the resonator to the lost energy on the length of the resonator is equal to TR=6.6>N2/3. In this case, the longitudinal fields $E_z$ (in stavolt/cm) after the 9th bunch, counted at 18 points along the resonator from z=0.1 cm to z=0.525 cm, are given in the table below and in Fig 14.

| z=0.1 | z=0.125 | z=0.15 | z=0.175 | z=0.2 | z=0.225 | z=0.25 | z=0.275 | z=0.3 |
|---|---|---|---|---|---|---|---|---|
| $E_{dec}$= 9.1×10$^5$ | $E_{dec}$= 1.9×10$^6$ | $E_{dec}$= 4.4×10$^5$ | $E_{dec}$= 2.7×10$^4$ | $E_{dec}$= 5.4×10$^5$ | $E_{dec}$= -1.6×10$^6$ | $E_{dec}$= -1.4×10$^6$ | $E_{dec}$= 7.2×10$^5$ | $E_{dec}$= 4.0×10^5 |
| $E_{ac}$= -3.5×10$^5$ | $E_{ac}$= -1.9×10$^5$ | $E_{ac}$= -3.6×10$^6$ | $E_{ac}$= -1.2×10$^6$ | $E_{ac}$= -3.4×10$^6$ | $E_{ac}$= -3.4×10$^6$ | $E_{ac}$= -5.5×10$^6$ | $E_{ac}$= -3.1×10$^6$ | $E_{ac}$= -3.3×10$^6$ |
| $E_{ac}/E_{dec}$ =0.38 | $E_{ac}/E_{dec}$ =0.10 | $E_{ac}/E_{dec}$ =8.3 | $E_{ac}/E_{dec}$ =44.7 | $E_{ac}/E_{dec}$ =6.6 | $E_{ac}/E_{dec}$ =-2.1 | $E_{ac}/E_{dec}$ =-4.1 | $E_{ac}/E_{dec}$ =4.4 | $E_{ac}/E_{dec}$ =8.2 |
| z=0.325 | z=0.35 | z=0.375 | z=0.4 | z=0.425 | z=0.45 | z=0.475 | z=0.5 | z=0.525 |
| $E_{dec}$= 2.3×10$^5$ | $E_{dec}$= 6.6×10$^5$ | $E_{dec}$= 5.5×10$^5$ | $E_{dec}$= -3.8×10$^5$ | $E_{dec}$= 1.6×10$^6$ | $E_{dec}$= 7.8×10$^4$ | $E_{dec}$= 4.6×10$^5$ | $E_{dec}$= 1.3×10$^6$ | $E_{dec}$= 3.0×10$^6$ |
| $E_{ac}$= -8.8×10$^5$ | $E_{ac}$= -8.8×10$^5$ | $E_{ac}$=- 2.5×10$^6$ | $E_{ac}$= -4.5×10$^6$ | $E_{ac}$= 3.3×10$^5$ | $E_{ac}$=- 4.3×10$^6$ | $E_{ac}$=- 3.6×10$^6$ | $E_{ac}$=- 2.9×10$^6$ | $E_{ac}$=- 3.4×10$^4$ |
| $E_{ac}/E_{dec}$ =3.8 | $E_{ac}/E_{dec}$ =1.3 | $E_{ac}/E_{dec}$ =4.5 | $E_{ac}/E_{dec}$ =-11.9 | $E_{ac}/E_{dec}$ =-0.2 | $E_{ac}/E_{dec}$= 55.1 | $E_{ac}/E_{dec}$ =7.9 | $E_{ac}/E_{dec}$ =2.3 | $E_{ac}/E_{dec}$ =0.01 |

**Table 2**

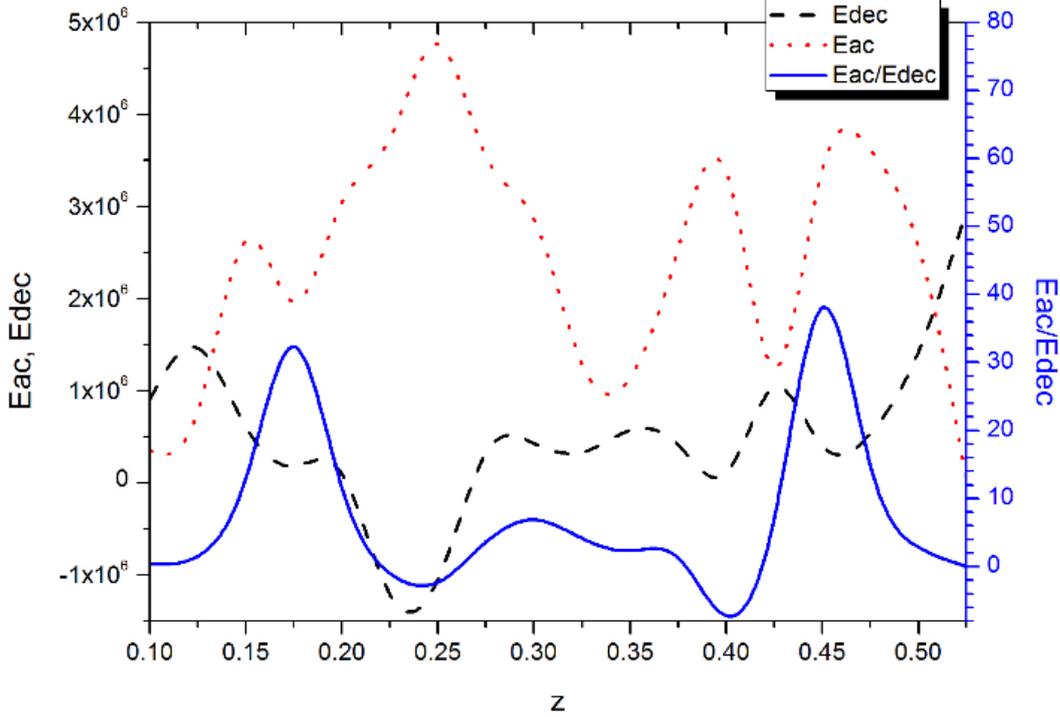

Fig. 14. The decelerating field (in stavolt/cm) inside the 9th bunch $E_{dec}$, $E_{ac}$=-$E_{acc}$, $E_{acc}$ is the accelerating field after the 9th bunch and their ratio $E_{ac}/E_{dec}$ along the resonator (in cm)

## CONCLUSIONS

So it has been shown that in the case of wakefield excitation in dielectric resonator by sequence of rectangular electron bunches, the charge of which is shaped according to linear law, the transformer ratio can achieve large value.

To increase the breakdown threshold in the resonator, it is advantageous to use a short pulse of the RF field. It has been shown that in the case of a short resonator, it is useful to use the case of a charge-profiled sequence of electron bunches, since in this case a large transformer ratio of the energy of the driver-bunches to the energy of accelerated bunches is achieved, as well as at the moment of injection of accelerated electron bunches at the injection boundary the back front of the pulse with accelerating phase of the field is appeared. And when positron bunches are accelerated it is useful to use case of resonant sequence of driver electron bunches. Then, at moments of injection of accelerated positrons bunches on the boundary of injection, the back front of the pulse with the accelerating phase of the field appears.

It has been shown that in order to reduce the effect of transient radiation on the transformer ratio, it is useful to use bunches of finite length.

As a result of the effect of the back wave of large amplitude on a small decelerating field, the driver bunch gets into the accelerating field at some points along the resonator.